\documentclass[fleqn,usenatbib]{mnras}
\usepackage{newtxtext,newtxmath}

\usepackage[T1]{fontenc}
\usepackage{ae,aecompl}

\usepackage{graphicx}	% Including figure files
\usepackage{amsmath}	% Advanced maths commands
\usepackage{amssymb} % Extra maths symbols

\usepackage{textcomp}
\usepackage{mathtools}
\usepackage{pict2e}
\usepackage[utf8x]{inputenc} 

\title[Simulations of the polarized radio sky]{Simulations of the polarized radio sky and predictions on the confusion limit in polarization for future radio surveys}

\author[F. Loi et al.]{
F. Loi$^{1,2}$\thanks{E-mail: francesca.loi@inaf.it},
M. Murgia$^{2}$,
F. Govoni$^{2}$,
V. Vacca$^{2} $,
I. Prandoni$^{3}$,
A. Bonafede$^{1,3,4}$,
\newauthor and L. Feretti$^{3}$.
\\
$^{1}$Dip. di Fisica e Astronomia, Universit\`a degli Studi Bologna, Viale Berti Pichat 6/2, I--40127 Bologna, Italy \\
$^{2}$INAF - Osservatorio Astronomico di Cagliari, Via della Scienza 5, I-09047 Selargius (CA), Italy \\
$^{3}$INAF - Istituto di Radioastronomia, Via Gobetti 101, I--40129 Bologna, Italy \\
$^{4}$ Hamburger Sternwarte, Universit\"at Hamburg, Gojenbergsweg 112, 21029, Hamburg, Germany.\\
}

\date{Accepted XXX. Received YYY; in original form ZZZ}

\pubyear{2018}
\begin{document}
\label{firstpage}
\pagerange{\pageref{firstpage}--\pageref{lastpage}}
\maketitle
% Abstract of the paper
\begin{abstract}
Numerical simulations offer the unique possibility to forecast the results of surveys and targeted observations that will be performed with next generation instruments like the Square Kilometre Array.
In this paper, we investigate for the first time how future radio surveys in polarization will be affected by confusion noise. 
To do this, we produce 1.4\,GHz simulated full-Stokes images of the extra-galactic sky by modelling various discrete radio sources populations.
The results of our modelling are compared to data in the literature to check the reliability of our procedure. We also estimate the number of polarized sources detectable by future surveys.
Finally, from the simulated images we evaluate the confusion limits in $I$, $Q$, and $U$ Stokes parameters, giving analytical formulas of their behaviour as a function of the angular resolution. 
\end{abstract}

 \begin{keywords}
polarization $-$ radio continuum: galaxies $-$ methods: numerical
\end{keywords}

\section{Introduction}
The capabilities of forthcoming radio telescopes, such as the Square Kilometre Array\footnote{https://www.skatelescope.org/} (SKA) and its precursors, will allow us to study the sky with an unprecedented detail and they will dramatically improve our knowledge of the radio Universe.

One of the main advantages of next generation radio-continuum surveys will be the possibility to study the faint signals coming from the most distant regions of the Universe over large field of views both in total intensity and in polarization. This is extremely important for a number of scientific applications, from the study of the physical and evolutionary properties of different classes of radio sources, to the investigation of the cosmic magnetism.\\
Concerning the first topic, important steps forward are expected from the radio continuum surveys that will be carried out with the SKA precursors: the Evolutionary Map of the Universe \citep[EMU,][]{norris} planned with the Australian Square Kilometre Array Pathfinder (ASKAP), the MeerKAT International GHz Tiered Extragalactic Exploration (MIGHTEE) survey \citep{jarvis}, the Westerbork Synthesis Radio Telescope (WSRT) Apertif \citep{norris13}, and the Very Large Array (VLA) Sky Survey (VLASS) \citep{lacy}. For a detailed discussion of the scientific expectations of the SKA for continuum science we refer to \citet{prandoni}.\\
Regarding cosmic magnetism, the origin and the evolution of large scale magnetic fields have not yet been established, despite many observational and numerical simulation-based efforts.
To determine the characteristics of large scale magnetic fields in galaxy clusters, one can analyse the Faraday rotation which affects every linearly polarized signal (the one from a background radio source) passing through a magnetised plasma (the intra-cluster medium) \citep[see the reviews on the determination of cluster magnetic fields of][]{cartay,govoni04}.
The Faraday rotation of extra-galactic radio sources can also be used to evaluate the Galactic magnetic field. \citet{taylor} have used the NRAO VLA Sky Survey \citep[NVSS,][]{condon98} at 1.4\,GHz to produce a rotation measure (RM) Grid which has an average of 1 polarized source per square degree. These data have been used by \citet{oppermann} to produce a reconstruction of the Galactic foreground Faraday rotation.
Since the sensitivity of future radio surveys will significantly improve, it will be possible to realise a denser RM Grid.
In this framework an important step forward will be represented by the polarization Sky Survey of the Universe’s Magnetism \citep[POSSUM,][]{gaensler}, that will be carried out with ASKAP. POSSUM will make use of the same full Stokes observations dedicated to EMU, and therefore will share the same observational parameters (rms noise $\sim$10\,$\muup$Jy beam$^{-1}$, 10$^{\prime\prime}$ of resolution). While EMU will produce total intensity images, POSSUM will use the data to extract polarization and RM information producing a RM grid of approximately 25 polarized sources per square degree. 
In its first phase of implementation, the mid frequency element of SKA (SKA1-MID) is expected to reach an average of 230$-45$0 RMs per square degree at the sensitivity of 4\,${\rm \muup Jy\,beam^{-1}}$ with a resolution of 2${\rm^{\prime\prime}}$ \citep{melanie}. 

Radio observations performed with next generation radio telescopes would be sensitive enough to be potentially limited by confusion rather than thermal noise.
Confusion is an additional noise term due to the presence of background unresolved sources whose signal enters into the synthesised beam of the telescope. 
It is therefore clear that the larger the beam, the higher the confusion noise term.
In total intensity the behaviour of the confusion noise as a function of angular resolution have been extensively studied in the literature \citep[see][]{condon74,condon2002,condon2012}.
On the other hand, confusion noise has never been investigated in polarization, as the polarized signal from background radio sources is typically a factor 10-100 lower than the total intensity signal, and it has never been an issue in existing polarization surveys. 
However, this may be not true for the upcoming generation of extremely deep radio surveys, that may be confusion limited also in polarization.

This work aims at estimating the confusion noise in polarization at 1.4\,GHz.  
Generally, the existing studies in the literature make use of analytical formulas to estimate confusion at a given angular resolution. Such formulas are based on extrapolations of the observed source counts, assumed to follow a power law with slope and normalisation depending on observing frequency and depth.\\
In this work, we use a different approach, that relies on end to end simulations.
We simulate $I$, $Q$, and $U$ Stokes images of a synthetic population of discrete radio sources distributed over cosmological distances and we analyse them 
to evaluate the confusion limit at different angular resolutions both in total intensity and in polarization.  \\
The paper is organised as follows: 
in Section 2, we describe the models and the procedure adopted to produce spectro-polarimetric images of a population of discrete radio source; in Section 3, we show the comparison with data at 1.4 GHz, giving our expectation on the number of polarized source that future surveys could detect; in Section 4, we present the confusion limit in $I$, $Q$, and $U$ Stokes parameters and the analytical formulas that describe its behaviour as a function of the angular resolution; in Section 5, we discuss about the applicability of the obtained results. Finally, the conclusions are drawn in Section 6. 
Throughout the paper, we adopt a ${\rm\Lambda}$CDM cosmology with ${\rm H_0=71\,km \, s^{-1} Mpc^{-1}}$, ${\rm\Omega_m=0.27}$, and ${\rm\Omega_{\Lambda}=0.73}$.

%%%%%%%%%%%%%%%%%%%%%%%%%%%%%%%%%%%%%%%%%%%%%%%%%%%%%%%%%%%%%%
\section{Modelling the radio sky}
\begin{figure}
    \centering
    \includegraphics[width=0.47\textwidth]{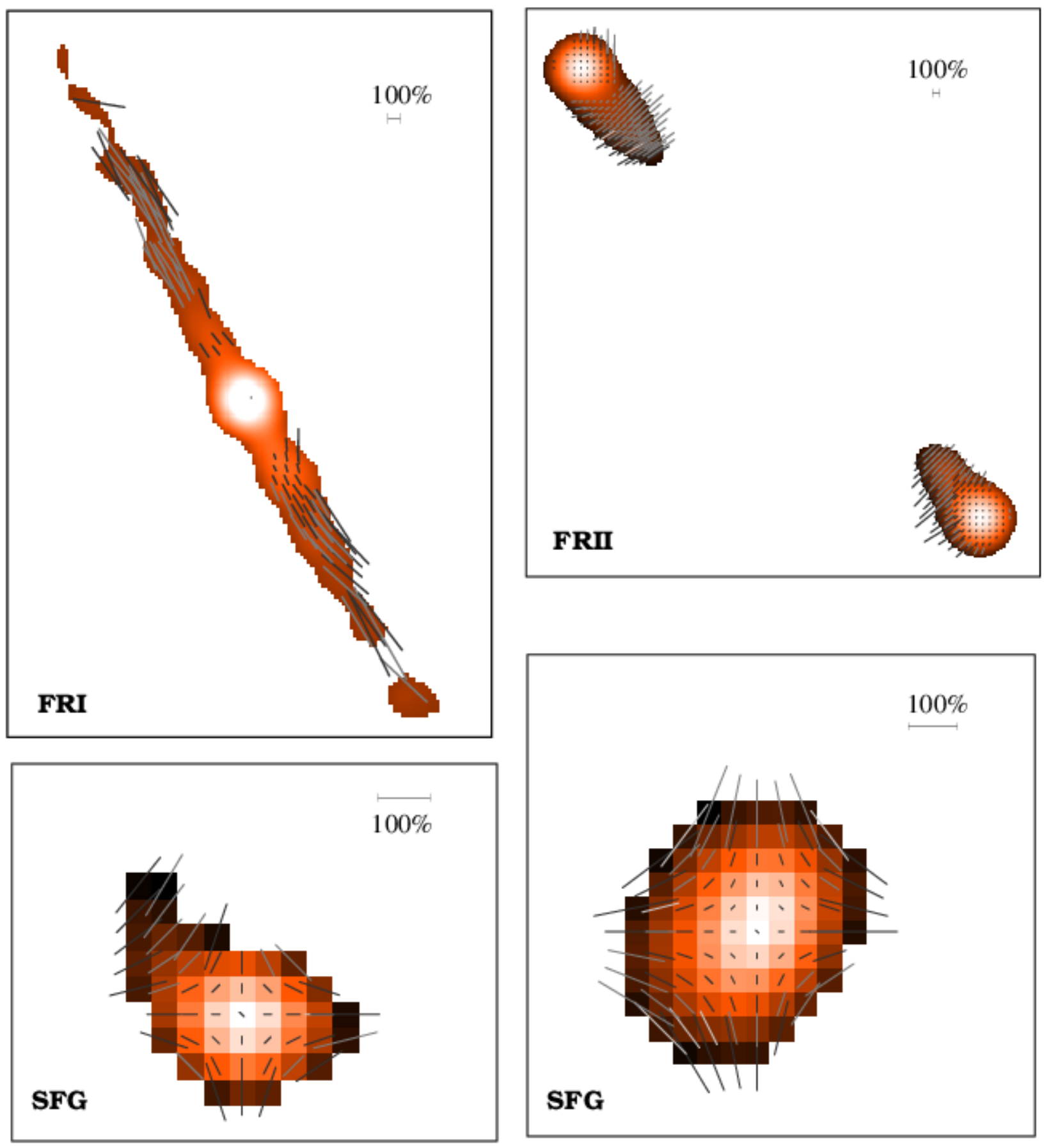}
    \caption{Models of radio galaxies where the colour represents the total intensity surface brightness distribution (normalised to one) and the vectors the intrinsic polarization strength and orientation. On the top, from left to right, we can see models of respectively Fanaroff-Riley (FR) type I and type II \citep{fr}, while in the bottom we show different models of SFGs.}
 \label{fig:model}
\end{figure}

\begin{figure}
    \centering
    \includegraphics[width=0.49\textwidth]{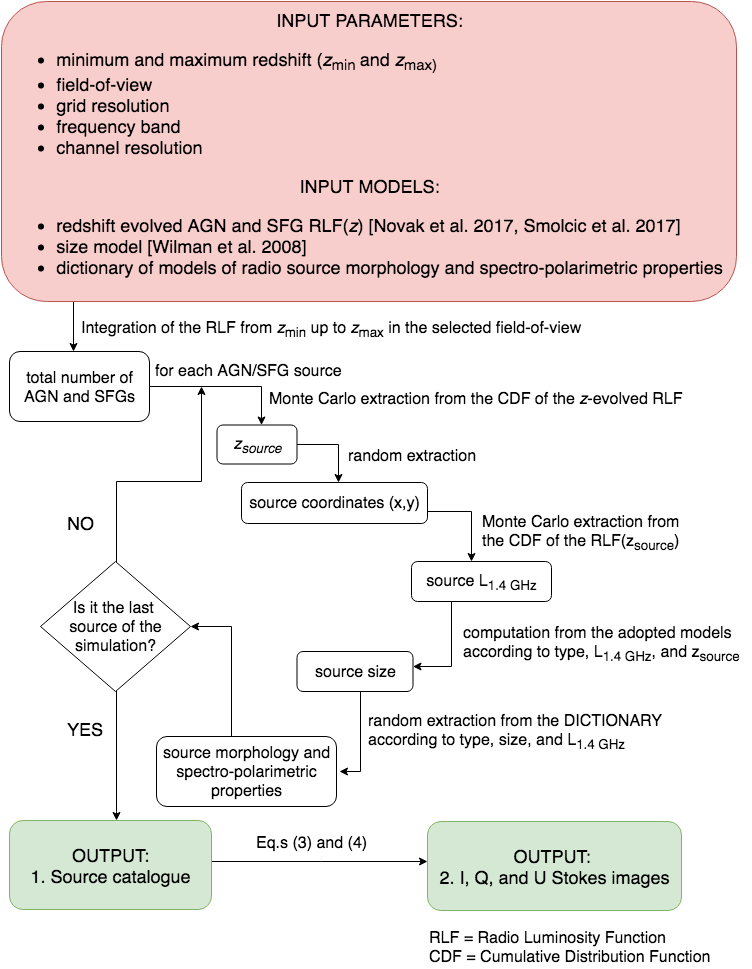}
    \caption{A flow chart of the procedure used to obtain simulated radio images.}
    \label{fig:flow}
\end{figure}

For this project, we make use of the FARADAY software package \citep{murgia04} which has been further developed to reproduce the polarized emission of a population of discrete radio sources.\\
As a first step, we produce a simulated catalogue of radio sources, generated by implementing recent determinations of the radio luminosity function (RLF) for the two main classes of objects dominating the faint radio sky: star forming galaxies (SFG) and Active Galactic Nuclei (AGN). 
The resulting catalogue contains all the discrete radio sources inside the "conical" portion of Universe whose angular aperture is set by the chosen field-of-view and whose depth extends from redshift $z$=0 up to a given $z$=$z_{\textrm{max}}$. \\
It is worth mentioning that simulated radio source catalogues already exist in the literature.
An example is the one produced by \citet{wilman08} which with a semi-empirical approach, starting from radio luminosity functions, simulates the radio continuum (total intensity) and HI emission of several radio source populations. Assuming a luminosity dependence for the fractional polarization, \citet{osullivan} realised a simulated polarized image based on the radio source catalogue of \citet{wilman08}.
Very recently a new simulated catalogue was produced \citep[T-RECS;][]{bonaldi} based on cosmological dark matter simulations to reproduce the clustering of sources and it models the radio sky both in total intensity and polarization with updated information on radio sources.
Our simulation,
like the above simulations, aims at giving useful information for the advent of the SKA.
Similarly, it is based on cosmological radio luminosity functions integrated over cosmological volumes but the models adopted to reproduce the characteristics of the radio sources and also the procedure are in general different. In addition, alternatively to the previous works, we use observed high-quality images of extended radio sources to reproduce the morphology and the spectro-polarimetric properties of the simulated radio sources. This is especially important as these simulations will be used to study magnetic fields in galaxy clusters (Loi et al. in prep.).\\ 
For each simulated radio source, our catalogue lists the following parameters:
\begin{itemize}
    \item {\it type}, in principle we can classify our sources in several sub-classes, radio-loud or radio-quiet AGN, SFG, quasar etc. Following \citet{novak} and \citet{smolcic} we consider two main families depending on the mechanism that triggers the radio emission: SFG and AGN;
    \item \textit{redshift}, z;
    \item \textit{size}, we used the relations adopted by \citet{wilman08} for radio-loud AGN and SFGs. The size model are redshift dependent and in particular the SFG size depends also on luminosity;
    \item \textit{luminosity at 1.4\,GHz}, we extract this information from the RLFs of \citet{novak} and \citet{smolcic} for the SFGs and AGN respectively, based on the results of the VLA$-$COSMOS 3\,GHz Large Project \citep{smolcic}, extrapolated to 1.4\,GHz assuming the spectral index derived in combination with the the VLA$-$COSMOS 1.4 GHz Large and Deep Projects \citep{schi1,schi2,schi3};
    \item \textit{coordinates}, (x,y); 
    \item \textit{morphology and spectro-polarimetry properties}, we select a model of radio source from a dictionary depending on its luminosity and type. Each model of this dictionary consists of four 1.4\,GHz images:
    \begin{itemize}
        \item the surface brightness ${ I}_{\nu}$ in total intensity;
        \item the spectral index distribution $\alpha$ determined by assuming that the flux density ${S}_{\nu}$ at a frequency ${\nu}$ is ${S}_{\nu}\propto\nu^{-\alpha}$;
        \item the fractional polarization defined as the ratio between the polarized intensity and the total intensity $ FPOL=P/I$;
        \item the intrinsic polarization angle which is defined with respect to the $Q$, and $U$ Stokes parameters as:
        \begin{equation}
                  \Psi_0=0.5 \cdot \arctan{U/Q}.
        \label{eq:psi}
        \end{equation}
    \end{itemize}
    The images of the dictionary are real high-quality high-resolution images performed at high-frequency. In particular, we used VLA images at C and X bands at arcsecond resolution so that the polarization properties can be considered very close to the intrinsic values. Some examples of models are shown in Fig. \ref{fig:model}, where the colour represents the total intensity surface brightness (normalised to one) and the vectors the intrinsic polarization strength and orientation. For the AGN class we consider sources with two different morphology: Fanaroff-Riley (FR) type I and type II \citet{fr}. For the SFG class we use images of spiral galaxies.
\end{itemize}    
From an operative point of view, the generation of the catalogue is generally based on a Monte Carlo extraction from the corresponding cumulative distribution functions of the models. A flow chart of the adopted procedure is shown in Fig. \ref{fig:flow}.\\
As a first step, we set the maximum redshift up to which we populate the simulated portion of the Universe. 
We split the slice into sub-volumes of ${\rm\Delta}z$=0.01 in width.
We perform the integral of the AGN and SFG RLFs throughout the solid angle of the simulated observation sub-volume by sub-volume : the result is the total number of ``cosmological'' AGN and SFGs respectively. As a maximum redshift we set $z_{\rm max}$=6, since the adopted RLFs sample AGN and SFGs up to a redshift of $z$=5.7 and $z$=5.5 respectively.
The radio source redshift is assigned through a Monte Carlo extraction from the cumulative distribution computed for each specific type from the corresponding RLF evolution.
We populate each sub-volume by randomly extracting the coordinates. The luminosity is assigned from the corresponding cumulative distribution function based on the evolved RLF at the redshift of the radio source.
We compute the source size taking into account the redshift  and also the luminosity in the case of SFGs. 
A model for the radio source is extracted from the dictionary according to the luminosity and type. The surface brightness distribution at a given frequency is re-scaled such that the luminosity:
\begin{equation}
  L_{\nu}=\int_{\Sigma} I_{\nu} (x,y)\, dx dy 
  \label{eq:l}
\end{equation}
matches the one assigned to the source, where the integral is performed over the radio source area ${\rm \Sigma}$.

Once we obtain our simulated catalogue of radio sources, we set the frequency bandwidth and channel resolution, and we used FARADAY to generate a spectral-polarimetric cube for each source and for each of the  $I$, $Q$, and $U$ Stokes parameters. In this process, the algorithm considers the correct spectral index for each pixel according to the catalogue. Indeed, the observed surface brightness at a given pixel of coordinates ($x,y$) depends on the redshift $z$ and on the spectral index at the corresponding coordinates ${\rm \alpha}$($x,y$):
 \begin{equation}
  I_{\nu} (x,y) =\frac{L_{\nu}}{A} \frac{1}{(1+z)^{3+\alpha(x,y)}}
 \end{equation}
 where $A$ is the pixel area.\\
By multiplying the surface brightness and the fractional degree polarization maps, we obtain the intrinsic polarized intensity of the selected radio galaxy. The radio sources which constitute our dictionary are not enough to represent the level of polarization statistically observed and reported in the literature. This is why we decided to re-scale the fractional polarization images in such way that the AGN and the SFGs can assume values between 0$-$10\% and 0$-$5\% respectively as observations of statistical samples suggest \citep{hales}. 
The $Q$, and $U$ Stokes parameters are computed by combining Eq. \ref{eq:psi} and ${ p_{\nu}= \sqrt{ U_{\nu}^2+Q_{\nu}^2}}$, where $\rm p_{\nu}$ is the polarized intensity at a given frequency $\rm \nu$:
\begin{eqnarray}
    Q_{\nu} & = &  \frac{p_{\nu}}{\sqrt{\tan^2{2\Psi_0}+1}} \nonumber \\
    U_{\nu} & = &  \frac{p_{\nu}\tan{2\Psi_0}}{\sqrt{\tan^2{2\Psi_0}+1}}
\end{eqnarray}
We neglect the effect of the Galactic Rotation Measure (RM) and we assume that no other magnetised plasma is present in the simulated portion of Universe. Otherwise, the observed polarized intensity would not be equal to the intrinsic one and we should compute the U and Q Stokes parameters starting from the polarization angle $\Psi$ defined as:
\begin{equation}
 { \Psi=\Psi_0+\rm \lambda^2 \cdot \rm\phi(l)},
 \label{eq:faraday}
\end{equation}
 where the $\rm \phi(l)$ is the Faraday depth defined as the integral performed over the length l (in kpc) of the crossed magneto-ionic plasma of the line-of-sight parallel component of the magnetic field $\rm B_{||}$ (in $\muup$G) times the thermal density $\rm n_e$ (in cm$^{-3}$):
 \begin{equation}
 \phi(l)=812 \int_0^l  n_{\rm e} \,{\bf B} \cdot{\bf dl}=812 \int_0^l  n_{\rm e} \,{ B}_{||} {dl} \quad {\rm [rad m^{-2}].}
 \label{eq:rm}
\end{equation}

%%%%%%%%%%%%%%%%%%%%%%%%%%%%%%%%%%%%%%%%%%%%%%%%%%%%%%%%%%%%%%%
\section{Comparison with data: total intensity and polarization source counts}
\label{sect:comp}
To test the reliability of our simulations, we compare our results with total intensity and polarization source counts available from the literature.\\
\begin{figure}\centering
 \includegraphics[width=0.47\textwidth]{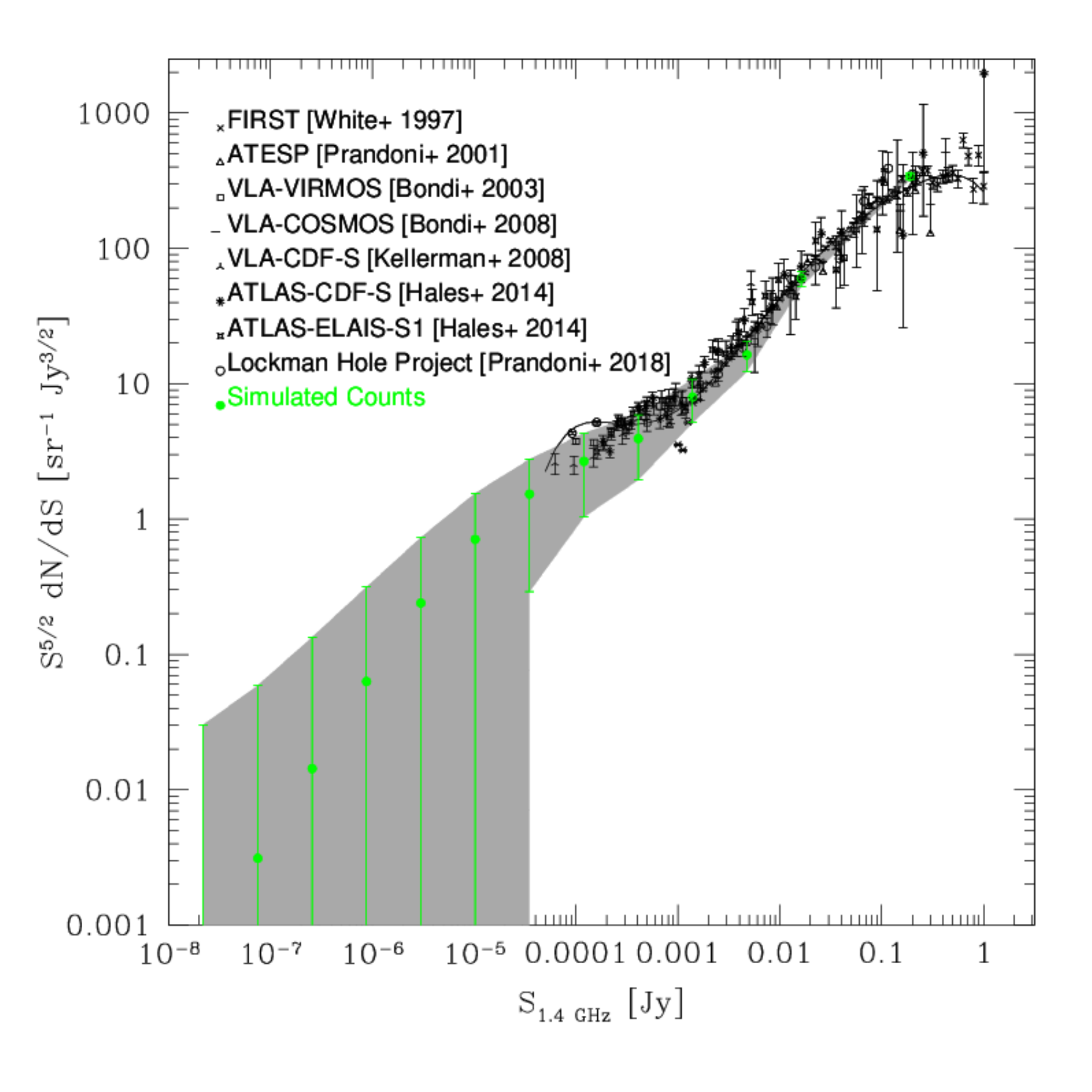}
\caption{Euclidean-normalised source counts at 1.4\,GHz: the black points represent the data \citep{white, atesp, bondi, bondi2008, kellermann, hales, prandoni18} while the green ones show the values obtained from the simulation of this work.}
\label{fig:cnt_I} 
\end{figure}
In Fig. \ref{fig:cnt_I}, we plot the 1.4\,GHz differential source counts of our simulated radio source population together with those estimated from surveys sampling at a wide flux density range, from $\sim$60\,$\muup$Jy up to 1\,Jy.
The counts are Euclidean normalised\footnote{the source counts are normalised with respect to an Euclidean Universe, where the number N of sources depends on their flux densities S as ${\rm \log N = S^{2/5} \log S}$} and the data refers to large-scale (> few square degree) 1.4\,GHz surveys \citep{white, atesp, bondi, bondi2008, kellermann, hales, prandoni18}. 
The flux density is evaluated taking into account the $k$-correction:
\begin{equation}
     S_{\nu}=\frac{L_{\nu}}{4 \pi D_{\rm L}^2} \cdot (1+z)^{1-\alpha}
\end{equation}
where $D_{\rm L}$ is the luminosity distance.
As shown in the plot, the simulated differential counts (green points) are in agreement with data.
This simulation can be used to predict the radio sky at sub$-\muup$Jy fluxes, that will be accessible with the next generation radio telescopes like the SKA over large field-of-view.
\begin{figure}\centering
 \includegraphics[width=0.47\textwidth]{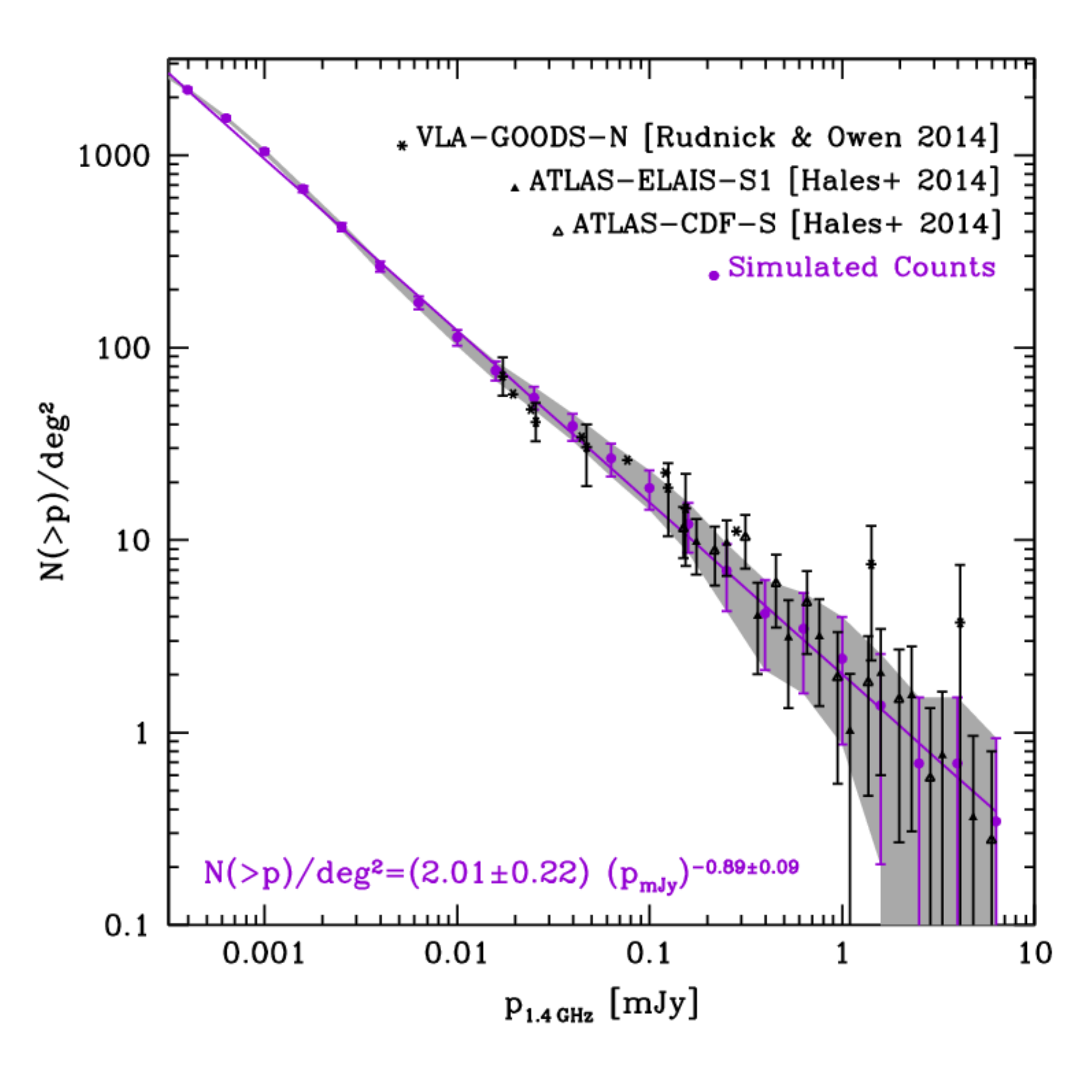}
 \caption{Cumulative counts of polarized sources as a function of the polarized flux density. The black points refer to the data of the GOODS-N field \citep{rudnick} and of the ATLAS data release 2 \citep{hales} while the red points represent the results obtained from the simulation of this work. The best-fit equation for the simulated counts is reported in the bottom left corner and it is represented as a solid purple line.}
 \label{fig:cfr_rudnick}
\end{figure}\\
In Fig. \ref{fig:cfr_rudnick}, we show the 1.4\,GHz cumulative counts of polarized sources as a function of the polarized source flux density ${\rm p_{\nu}}$ in mJy. 
The black points are 1.4\,GHz data \citep{hales,rudnick} which cover the range between ${\rm \sim 16\,\muup Jy}$ and ${\rm\sim60\,mJy}$ while the purple points are the cumulative counts obtained from our simulation. 
The error bars of the cumulative source counts $\sigma_N$ are the poissonian uncertainties. Even in this case, the agreement between data and simulation is remarkable.\\
We observe that the cumulative source counts as a function of the polarized flux density can be well described by a power-law:
\begin{equation}
    N(>p)/{\rm deg^2}=N_0 \cdot \left (\frac{p}{\rm mJy}\right )^{\gamma},
\end{equation}
which turns to be a linear function in the log-log space:
\begin{equation}
    y=A \cdot x + B,
\end{equation}
where $y=\log(N(>p)/{\rm deg^2})$, $x=\log(p)$, $B=\log(N_0)$, and $A=\gamma$. With the least squares method, we fit the cumulative counts obtaining the following relation:
 \begin{equation}
     N(>p)/{\rm deg^2}=(2.01\pm 0.22) \cdot \left (\frac{p_{\rm 1.4\,GHz}}{\rm mJy}\right )^{(-0.89\pm 0.09)},
    \label{eq:p}
\end{equation}
represented with a purple line in Fig. \ref{fig:cfr_rudnick}. In our fitting, we take into account the uncertainties on the measurements as:
\begin{equation}
    \sigma_y=\frac{1}{N(>p)/{\rm deg^2}}\cdot \ln{10}\cdot \sigma_N
\end{equation}
The errors associated to the parameters $N_0$ and $\gamma$ are then $\sigma_{N_0}=10^B \cdot \ln{10}\cdot \sigma_B$ and $\sigma_{\gamma}=\sigma_A$.\\
Also in this case, our simulation can be used to investigate radio source populations with polarized flux density lower than the limit of current observations.
In particular, Table \ref{tab:table} reports our expectations in terms of polarized source numbers and densities for several radio continuum polarization surveys.
From left to right, each column shows the survey name, 
the sensitivity level in polarization $\sigma_p$ in $\muup$Jy at 1.4\,GHz, the expected number of sources per square degree with polarized intensity higher than 3$\sigma_p$, the field of view of the survey in square degree, and the number of sources that each survey would detect. The number and the number density per square degree have been computed from Eq. \ref{eq:p}.
\begin{table}
	\centering
	\caption{From left to right, each column shows the survey name, 
the sensitivity level in polarization $\sigma_p$ in $\muup$Jy at 1.4\,GHz, the expected number of sources per square degree with polarized intensity higher than 3$\sigma_p$, the field of view of the survey in square degree, and the number of sources that each survey would detect. The number and the number density per square degree have been computed from Eq. \ref{eq:p}.}
	\label{tab:table}
	\begin{tabular}{l c r r r} % four columns, alignment for each
		\hline
		Survey & $\sigma_p$[$\muup$Jy] & N/deg$^2$ & FoV[deg$^2$] & N[$\times10^3$] \\
		\hline
		VLASS & 89 & 7 & 33885 & 220\\
		Apertif & 10 & 46 & 3500 & 159\\
		POSSUM & 7 & 63 & 30000 & 1877\\
		MIGHTEE & 0.7 & 486 & 20 & 10\\
		SKA1-MID all-sky & 2.8 & 141 & 31000 & 4385 \\
		\hspace{1.2cm} wide & 0.7 & 486 & 1000 & 486 \\
		\hspace{1.2cm} deep & 0.14 & 2034 & 30 & 61 \\
		\hspace{1.2cm} ultra-deep & 0.035 & 6987 & 1 & 7\\
		\hline
	\end{tabular}
\end{table}

%%%%%%%%%%%%%%%%%%%%%%%%%%%%%%%%%%%%%%%%%%%%%%%%%%%%%%%%%%%%
\section{The confusion limit in total intensity, Q, and U Stokes parameters}
The possibility to simulate all the radio sources that are present in a given field-of-view let us explore the effect of the confusion noise, which is due to the faint unresolved radio sources whose signal enters in the beam of the telescope. 
While we can reduce thermal noise by increasing the exposure time, confusion is a physical limit that we cannot overcome for a fixed maximum baseline length and it is important to have an accurate estimate of its statistical properties. 

Here, we simulate the full-Stokes parameters  at 1.4\,GHz of a radio source population in a computational grid corresponding to $\sim$\,0.72\,deg$^2$ with a resolution of 1${\rm ^{\prime\prime}}$.
\begin{figure*}\centering
\includegraphics[scale=0.46]{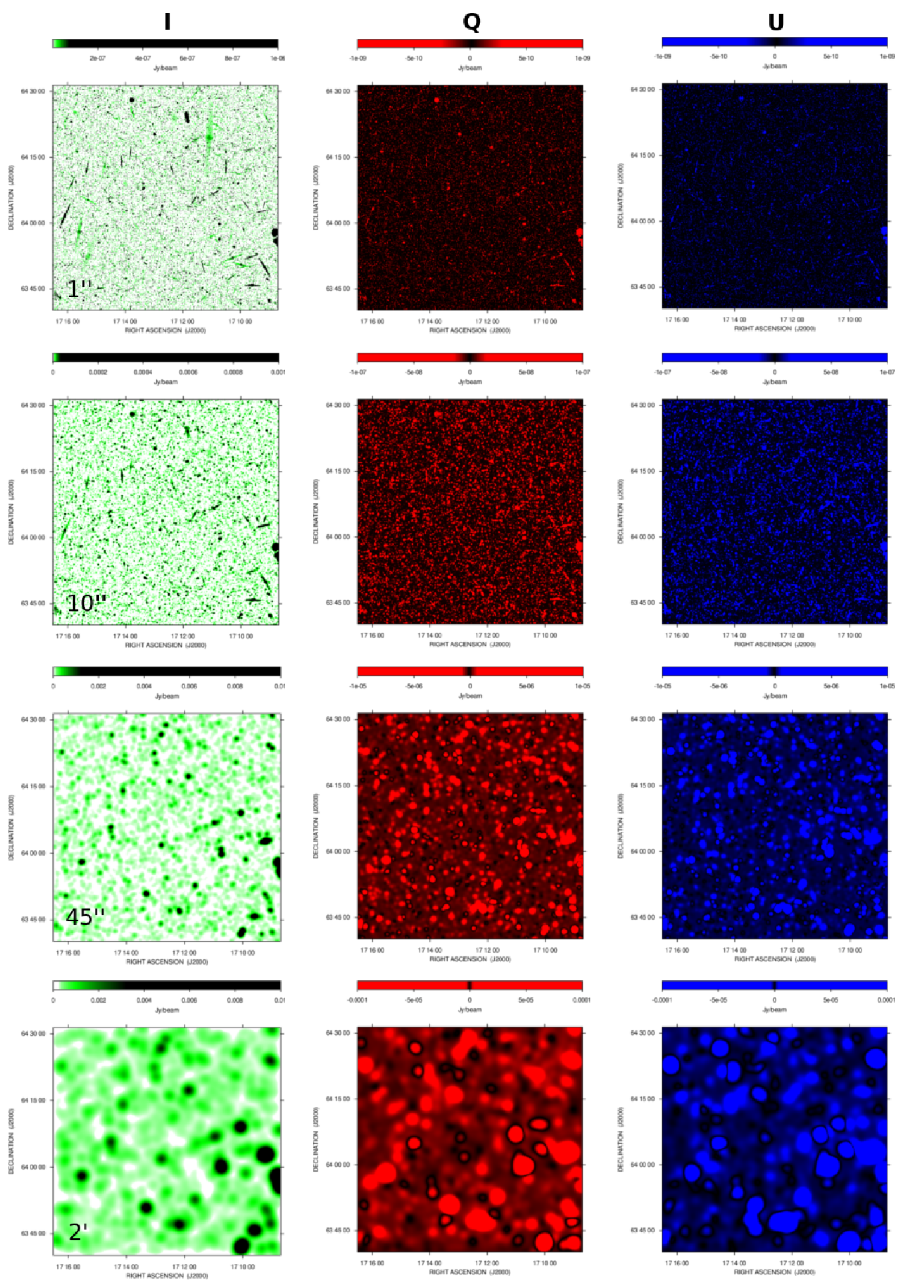}
\caption{From left to right, simulated images of the  $I$, $Q$, and $U$ Stokes parameters at 1.4\,GHz of radio galaxies in a field-of-view of ${\rm\sim 0.72\,deg^2}$ convolved with a beam FWHM equal to, from top to bottom, 1${\rm ^{\prime\prime}}$, 10${\rm ^{\prime\prime}}$, 45${\rm ^{\prime\prime}}$, and 2${\rm ^{\prime}}$ respectively.}
\label{fig:conf_img} 
\end{figure*}
The resulting images have been convolved with different beam sizes. In particular we consider beam Full-Width-at-Half-Maximum (FWHM) equal to 1${\rm ^{\prime\prime}}$, 2${\rm ^{\prime\prime}}$, 6${\rm ^{\prime\prime}}$, 10${\rm ^{\prime\prime}}$, 20${\rm ^{\prime\prime}}$, 45${\rm ^{\prime\prime}}$, 60${\rm ^{\prime\prime}}$, and 120${\rm ^{\prime\prime}}$. 
In Fig. \ref{fig:conf_img}, we show the resulting images at, from top to bottom, 1${\rm ^{\prime\prime}}$, 10${\rm ^{\prime\prime}}$, 45${\rm ^{\prime\prime}}$, and 120${\rm ^{\prime\prime}}$ beam FWHM. Columns, from left to right, show the  $I$, $Q$, and $U$ Stokes images respectively.
\begin{figure}
    \centering
    \includegraphics[width=0.47\textwidth]{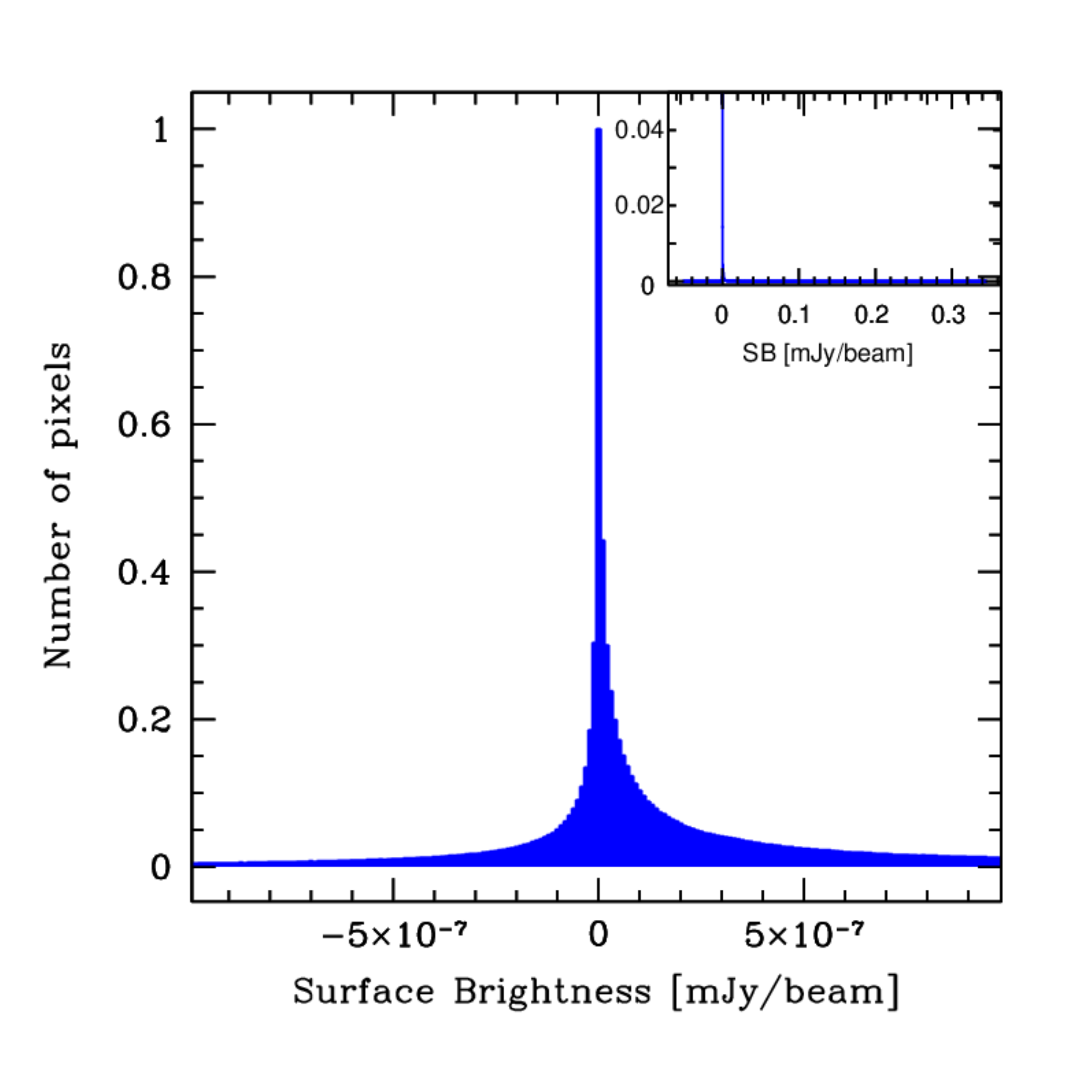}
    \caption{A zoom of the histogram of the surface brightness in mJy/beam as measured over the 1$^{\prime\prime}$ total intensity image of Fig. \ref{fig:conf_img}. The y-axis represents the number of the pixels at a given surface brightness normalised to 1. In the top right corner, a zoom out of the same histogram is reported to show the full range of values assumed by the distribution.}
   \label{fig:histo_conf}
\end{figure}\\
Starting from these images, we want to determine the confusion limits at different beam FWHM. In total intensity the spatial distribution of confusion sources over a large region of the sky forms a plateau characterised by a mean different from zero. However, this base level cannot be observed in the interferometric images due to the missing short space baselines in the $u-v$ plane. 
Thus, what we observe in general is the fluctuating component of the confusion. The distribution of these fluctuations is highly non-Gaussian but it presents a long tail at high flux densities due to bright sources. This is shown in Fig. \ref{fig:histo_conf}, where we plot the Stokes I surface brightness distribution obtained from the image of Fig. \ref{fig:conf_img} at 1$^{\prime\prime}$ of resolution. The y-axis represents the number of the pixels at a given surface brightness normalised to 1. In the top right corner, it is reported a zoom out of the same histogram to show the full range of surface brightness values assumed by the distribution. The long tail towards high surface brightness values is due to the presence of real sources.\\
In real images, the confusion is estimated from the probability distribution P(D) measured in a cold part of the sky, which corresponds to the distribution of the surface brightness image. The P(D) distribution is the convolution of the confusion due to the faint sources and the thermal noise which are independent of each other, so that the total observed variance ${\rm \sigma_o^2}$ is the sum of the variance due to the confusion noise ${\rm \sigma_c^2}$ and to the thermal noise ${\rm \sigma_n^2}$:
\begin{equation}
    \rm \sigma_o^2=\sigma_c^2+\sigma_n^2.
\end{equation}
To estimate the confusion, in general it is necessary to start from images where the ${\rm \sigma_c^2} \ll {\rm \sigma_n^2} $. The simulated images obtained in this work are not affected by any kind of noise except the confusion.
Therefore, to measure the confusion we could simply measure the rms from the simulated images. However, to be sure that we are not taking into account bright sources which should be distinguishable from the confusion, we measure the average and the rms with an iterative procedure.
For each image at a given beam resolution, we follow these steps: 
\begin{enumerate}
    \item we cover the image with boxes with sizes 10 times the beam FWHM;
    \item we evaluate the rms in every box by iterative clipping all the pixels having an intensity larger than 10${\rm \times}$rms, until convergence and no other pixels are excluded. In practice, we consider that the confusion noise is related only to the sources fainter than a signal-to-noise ratio S/N<10, where N is evaluated numerically by clipping the tail of the distribution as described above;
    \item we compute the confusion limit by averaging the rms values of the different boxes and its error as the square root of the standard deviation of the obtained mean divided by the number of boxes.
\end{enumerate}

\begin{figure}\centering
\includegraphics[width=0.47\textwidth]{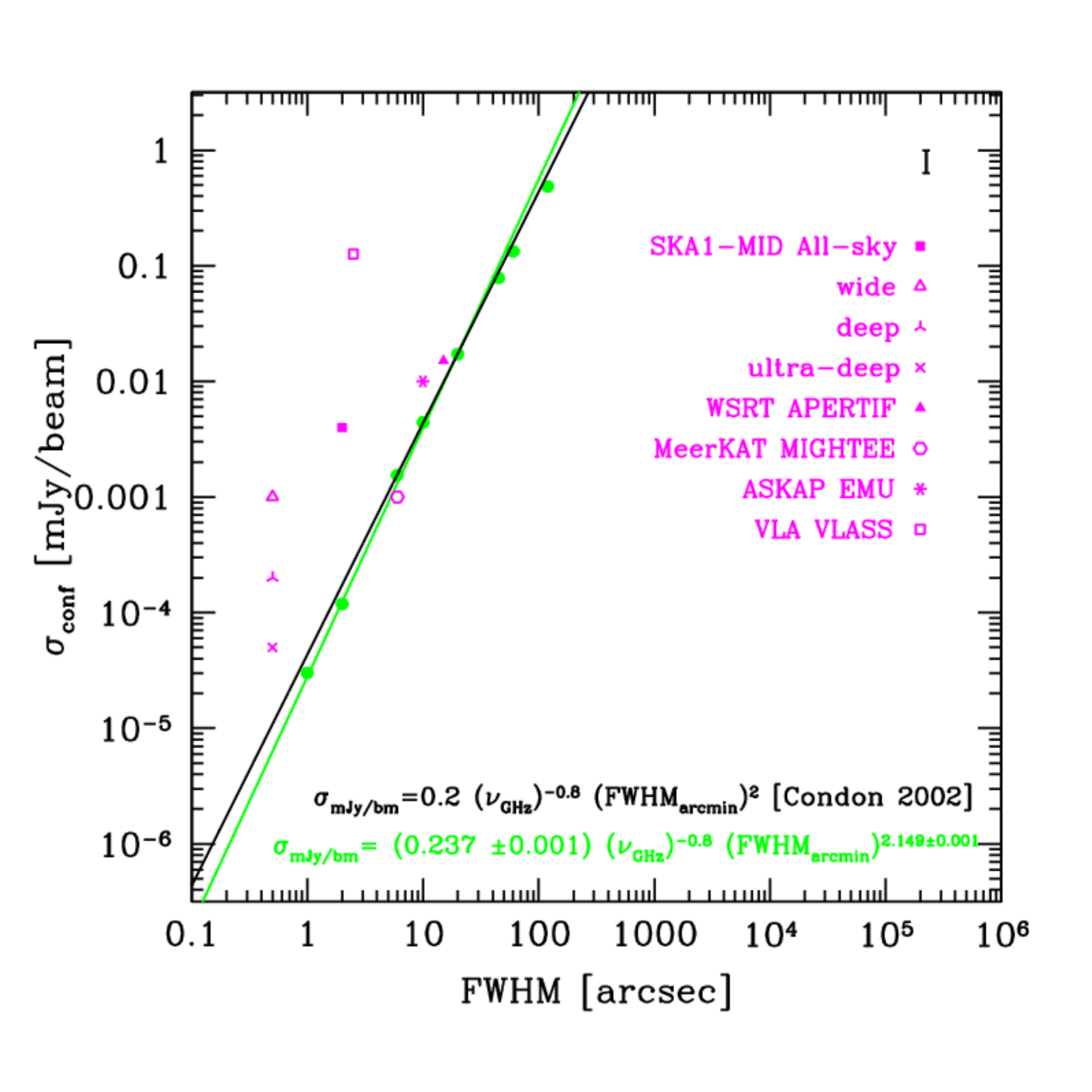}
\caption{The plot show the 1.4\,GHz confusion noise in total intensity calculated from the convolved images with respect to the FWHM as green dots fitted with the solid green line. The black line represents the formula proposed by \citet{condon2002} which is reported together with the fitted relation in the bottom left corner. We also plot in magenta the expected sensitivity of different surveys: the SKA1-MID all-sky, wide, deep, and ultra-deep surveys \citep{prandoni}, the WSRT Apertif survey \citep{norris13}, the ASKAP MIGHTEE survey \citep{jarvis}, the ASKAP EMU survey \citep{norris}, and the VLA VLASS \citep{lacy}.}
\label{fig:conf_I_plot} 
\end{figure}
The computed confusion limits in total intensity at different FWHM are plotted in Fig. \ref{fig:conf_I_plot}:
the measurements performed on the 1.4\,GHz simulated images are represented with green dots.\\
As for the case of the cumulative counts, we assume a power law behaviour for the confusion noise $\sigma=N_0 \cdot (FWHM)^\gamma$ as a function of the beam resolution, and we fit the results with the least squares method in the log-log space, where $y=\log(\sigma)$, $x=\log(FWHM)$, $B=\log(N_0)$, $A=\gamma$. We find the following relation:
\begin{equation}
      {\rm  \sigma_{1.4\,GHz}}^I = \rm {(0.1862\pm0.0009) \cdot \left (\frac{FWHM}{arcmin} \right)^{2.149\pm 0.001 } \,mJy/beam.}
        \label{eq:I}
\end{equation}
Assuming an average spectral index for the source population of $\alpha=0.8$ the previous relation can be written as:
\begin{equation}
         \sigma_{\nu}^I = (0.237\pm0.001) \cdot \left (\frac{\nu}{\rm GHz} \right )^{-\alpha}\cdot \left (\rm \frac{FWHM}{arcmin} \right )^{2.149\pm0.001} \,{\rm mJy/beam},
\end{equation}
where we consider $\left (\frac{\nu}{\rm GHz} \right )^{-\alpha}$ as a constant and therefore we simply divided the fitted parameter $N_0$ and its uncertainty by this constant.
Our results can be compared with the confusion noise expected on the basis of the formula provided by \citet{condon2002}: 
\begin{equation}
 \sigma_{\nu}^I=0.2 \cdot \left ( \frac{\nu}{\rm GHz} \right )^{-\alpha} \cdot \left ( \rm \frac{FWHM_{min} \cdot FWHM_{max}}{arcmin^2} \right)\, {\rm mJy/beam},
\end{equation}
where $\rm FWHM_{min}$ and $\rm FWHM_{max}$ are the minimum and the maximum beam FWHM. As reference, we trace this relation with a black line in Fig. \ref{fig:conf_I_plot} where we assume $\alpha$=0.8, $\nu$=1.4\,GHz and that $\rm FWHM_{min}=FWHM_{max}$.
We note that for what concerns the total intensity there is a remarkable agreement between the predictions of our simulations and the formula by \citet{condon2002} widely used in literature.
In the same Figure, we show in magenta the sensitivity levels foreseen for the same future surveys of Table \ref{tab:table}. As we can see, all the survey are very close to the confusion even at very high angular resolution, for example at 0.5$^{\prime\prime}$ resolution of the SKA1-MID ultra-deep survey, where the confusion noise is lower and it is possible to deeply explore the radio continuum sky.

No information in the literature has been reported so far about the confusion limit in $Q$, and $U$ Stokes parameters. The values measured in this work at different FWHM are plotted in Fig. \ref{fig:conf_UQ_plot}:
\begin{figure}\centering
\includegraphics[width=0.47\textwidth]{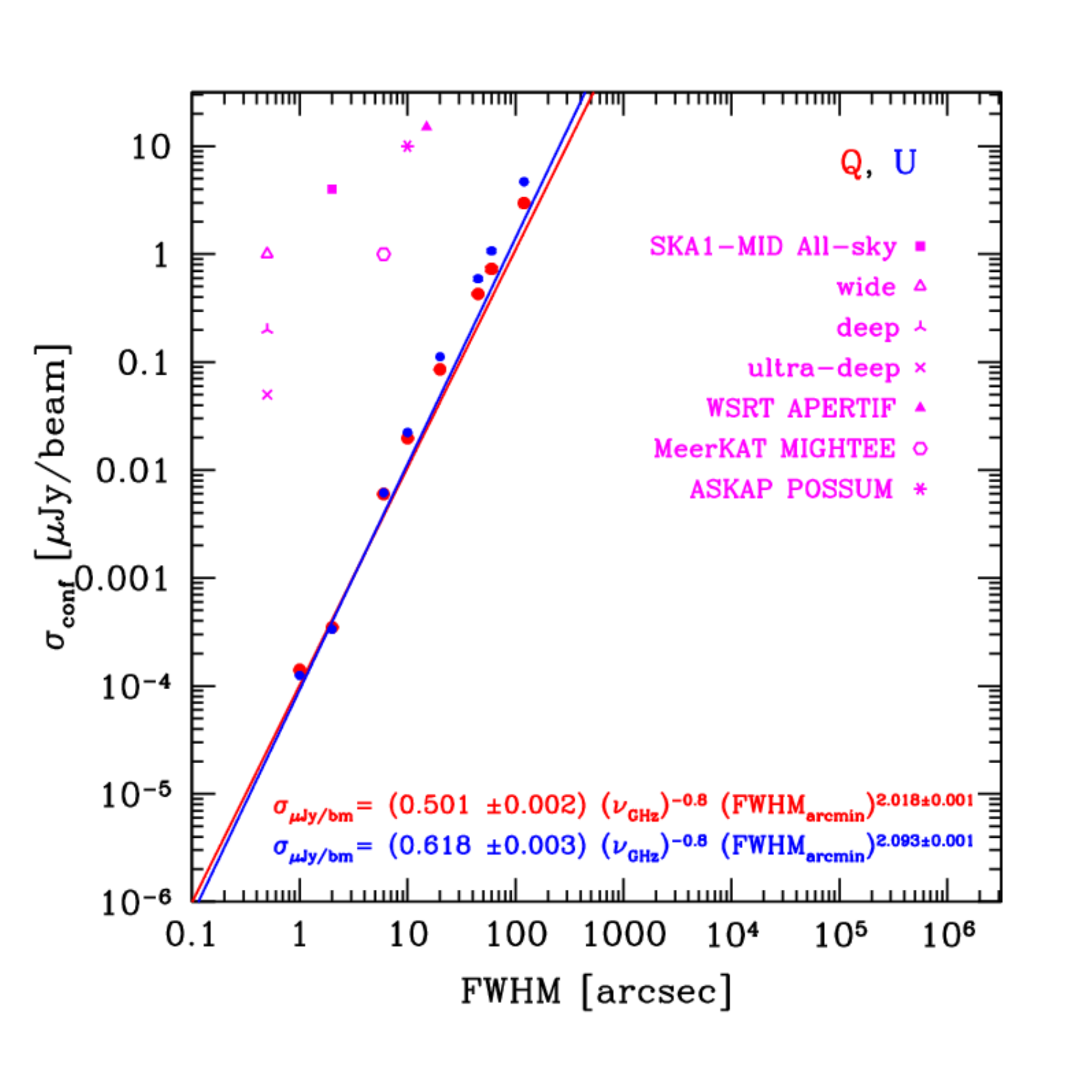}
\caption{The plot shows the 1.4\,GHz confusion noise versus the FWHM for the U (blue triangles and line) and Q (red dots and line) Stokes parameters. They have been fitted with the relations reported in the bottom left corner of the plot. We also plot in magenta the expected sensitivity of different surveys:  the all-sky, wide, deep, and ultra-deep SKA1-MID surveys \citep{prandoni}, the WSRT Apertif survey \citep{norris13}, the ASKAP MIGHTEE survey \citep{jarvis}, the ASKAP POSSUM \citep{gaensler} surveys.}
\label{fig:conf_UQ_plot} 
\end{figure}
the red and the blue solid line are respectively the fits of the 1.4\,GHz simulations of the $Q$, and $U$ confusion noise whose equations are indicated in the right bottom corner. 
We report in the following the best fit equations shown in the plot:
\begin{eqnarray}
 \sigma_{\rm 1.4\,GHz}^Q = (0.393\pm0.002) \cdot \left (\rm \frac{FWHM}{arcmin} \right )^{2.018\pm 0.001} \rm\, \muup Jy/beam \nonumber\\
 \sigma_{\rm1.4\,GHz}^U =(0.485\pm0.002) \cdot \left ( \rm \frac{FWHM}{arcmin} \right )^{2.093 \pm 0.001}\rm \, \muup Jy/beam
  \label{eq:U}
\end{eqnarray}
By assuming an average spectral index for the source population of $\alpha=0.8$, the previous relations can be written as: 
\begin{eqnarray}
   \sigma_{\nu}^Q=(0.501\pm 0.002) \cdot \left (\frac{\nu}{GHz} \right)^{-\alpha} \cdot \rm  \left ( \frac{FWHM}{arcmin} \right )^{2.018\pm 0.001}\rm \, \muup Jy/beam\nonumber \\
   \sigma_{\nu}^U=(0.618 \pm 0.003) \cdot \left (\frac{\nu}{GHz}\right)^{-\alpha} \cdot \rm \left  ( \frac{FWHM}{arcmin} \right )^{2.093\pm 0.001}\rm \, \muup Jy/beam
\end{eqnarray}
As expected the confusion limits of the U and Q Stokes parameters is lower than in total intensity, according to our simulation by a factor of $\sim$400. 
Concerning future surveys, we observe that in $Q$, and $U$ Stokes parameters the confusion limit is well below their sensitivity level, which has been reported in the same plot with magenta symbols. 
This represents an important result since, according to the modelling presented here, it means that with next generation telescopes we could perform very deep targeted observations in polarization without being limited by confusion noise.

%%%%%%%%%%%%%%%%%%%%%%%%%%%%%%%%%%%%%%%%%%%%%%%%%%%%%%%%%
\section{Applicability of the results}
\label{sect:app}
The simulations presented in this work aimed at determining the confusion limit in polarization as a function of angular resolution. \\
Our approach consists in a modelling of the discrete radio sources populating the Universe starting from their observed properties at 1.4\,GHz. 
Our investigation is based on a number of assumptions.
We discuss in the following the reasons behind and the possible limitations introduced by each of them.
\begin{enumerate}
    \item Frequency. At 1.4\,GHz the radio sky has been extensively studied down to $\muup$Jy flux levels, both in total intensity and in polarization. This enables us to compare our modelling with existing data in the literature and assess the reliability of our simulations. The results of the simulations at 1.4 GHz can be extrapolated to other frequencies by assuming an average spectral index for the various source populations. This approach has been followed by both \citet{wilman08} and \citet{bonaldi}, obtaining good results in reproducing observational trends, like source counts, etc. Nevertheless directly simulating the extra-galactic radio sky at lower and/or higher frequencies would certainly be the right approach to follow.
    \item Galactic foreground. We neglect the presence of a Galactic foreground. The effect of the Galactic RM is the rotation of the polarization plane of the signal, as shown in Eq. \ref{eq:faraday}. If we do not correct for the right value of Galactic RM the signal will be depolarized and measurements of the $Q$, and $U$ confusion limits would give values lower than what reported in this work. By applying techniques like the Rotation Measure Synthesis \citep{brent,burn}, it is possible to infer the Galactic RM value. Our results will correspond to the de-rotated $U$, and $Q$ Stokes images.
    \item Clustering. The simulated images used to estimate the confusion do not include clustering of sources. In other words, we are simulating a cold region of the sky, without galaxy clusters. The presence of source clustering would have the effect to create regions with different density of sources and likely a different distribution of the confusion. To evaluate the effect of clustering on confusion it is necessary to implement the clustering of sources along the filament of the cosmic web in our simulation and this is the goal of a future work. However, since our simulations agree with data (see Section \ref{sect:comp}), we are sure about the reliability of our results. If discontinuities in the number of sources can be clearly observed in images, our results would represent an average behaviour of the confusion between the higher and the lower density regions. It is worth noting that \citet{wilman08} include a clustering recipe in their simulations, but the results are questioned by radio source clustering analyses reported in the literature \citep[see e.g.][]{hale18}. \citet{bonaldi} also implemented source clustering in T-RECS, using a high-resolution cosmological simulation. 
    Issues that can be introduced by source over-densities is the possible presence of a magneto-ionic plasma in the inter-cluster medium and more generally in the filaments of the cosmic web. This will have the effect to depolarise the signal of background sources, resulting in a lower $Q$, and $U$ confusion limit. Provided that up to now the presence of magnetic fields in filaments is not firmly confirmed by observations \citep[but see][]{vacca18}, the magneto-hydro-dynamical simulations which explore this possibility suggest very weak magnetic fields in these structures \citep{vazza15} Therefore, the depolarization due to filaments should not have a significant impact on our estimates.
    However, the effect of source clustering mentioned here deserves dedicated studies and we consider them as a future prospect.
    \item Sidelobe contribution. An additional source of confusion, especially important in total intensity rather than in U and Q Stokes images, is due to the sidelobes of uncleaned sources lying outside the image. In the work presented here, we did not consider this contribution. This choice was made because we wanted to estimate the confusion noise due to the faint unresolved sources and compare it with the sensitivity foreseen for several surveys performed (or which are going to be performed) with different instruments. These instruments will be characterised by a different response, i.e. by a different (and sometimes still unknown) shape of primary beam, therefore the addition of the sidelobe contribution would make the results valid just for a particular instrument in a particular configuration.
    With this work, we give a first estimate of the confusion noise in polarization due to the faint unresolved sources. Thanks to this, we could focus on that instruments which seem capable to reach a thermal noise closed to the confusion value reported here and perform the analysis considering also the sidelobe contribution.
\end{enumerate}

\section{Conclusions}
In this work, we presented an original numerical approach developed to generate full-Stokes images of the radio sky. \\
We described the models and the procedure adopted to reproduce the discrete radio sources populating the Universe.\\
After that we successfully compared the results of our modelling and data from the literature concerning the differential source counts in total intensity and the cumulative source counts of polarized sources, we identified a simple functional relation between the number of polarized sources per square degree and the polarized flux density. From this relation, we computed the number of polarized sources that future surveys will detect, an useful information especially for cosmic magnetism investigations.
Finally, we evaluated the confusion limits in $I$, $Q$, and $U$ Stokes parameters at different beam resolution. Even in this case we found analytical formulas which describe the confusion limits as a function of the angular resolution. These formulas can be used as additional input for setting up observational strategies to maximise the impact of the next generation radio telescopes.

\section*{Acknowledgements}
We gratefully acknowledge the anonymous referee for the useful comments and suggestions.
FL and AB acknowledge financial support from the Italian Minister for Research and Education (MIUR), project FARE SMS, code R16RMPN87T. AB acknowledges financial support from the ERC-Stg DRANOEL, no 714245. IP acknowledges funding from the INAF PRIN-SKA 2017 project 1.05.01.88.04 (FORECaST). 
The trg computer cluster was funded by the Autonomous Region of Sardinia (RAS) using resources from the Regional Law 7 August 2007 n. 7 (year 2015) "Highly qualified human capital", in the context of the research project CRP 18 "General relativity tests with the Sardinia Radio Telescope" (P.I. of the project: Dr. Marta Burgay).

%%%%%%%%%%%%%%%%%%%%%%%%%%%%%%%%%%%%%%%%%%%%%%%%%%

%%%%%%%%%%%%%%%%%%%% REFERENCES %%%%%%%%%%%%%%%%%%

%%%%%%%%%%%%%%%%%%%%%%%%%%%%%%%%%%%%%%%%%%%%%%%%%%

\bsp	% typesetting comment
\label{lastpage}
\end{document}